\documentstyle[pra,twocolumn,aps,epsf]{revtex}

\begin{document}
\draft
\title{Hydrogen atom in a magnetic field: The quadrupole moment}

\author{Alexander Y. Potekhin\thanks{Electronic address:
 palex@astro.ioffe.rssi.ru}}
\address{Ioffe Physico-Technical Institute,
     194021 St.\ Petersburg, Russia}
\author{Alexander V. Turbiner\thanks{
On leave of absence from the Institute for Theoretical and
Experimental Physics,  Moscow 117259, Russia}}
\address{Laboratoire de Physique Th\'eorique, Universit\'e
Paris-Sud, F-91405, France}
\address{Instituto de Ciencias Nucleares,
UNAM, Apartado Postal 70-543,
      04510 M\'exico Distrido Federal, M\'exico}

\date{Received 8 January 2001}
\maketitle

\begin{abstract}

The quadrupole moment of the hydrogen atom in a magnetic field $B$
for field strengths from 0 to $4.414\times 10^{13}$~G is calculated
by two different methods. The first method is variational, and
based on a single trial function. The second method deals with a
solution of the Schr\"odinger equation in the form of a linear
combination of Landau orbitals.

\end{abstract}


\newcommand{\beq}{\begin{equation}}
\newcommand{\eeq}{\end{equation}}
\newcommand{\bea}{\begin{eqnarray}}
\newcommand{\eea}{\end{eqnarray}}
\newcommand{\req}[1]{Eq.~(\ref{#1})}
\newcommand{\mel}{m_e}
\newcommand{\dd}{{\rm\,d}}
\newcommand{\mc}[1]{\multicolumn{1}{c}{#1}}
\newcommand{\JPB}[1]{{J.\ Phys. B
} {\bf #1}}

\section{Introduction}

Plenty of works have been devoted to study of a hydrogen atom in a magnetic
field (see, e.g., Refs.\ \cite{CV77,Garstang:1977,Ruder:1994})
and this problem was among the first ever studied in quantum
mechanics. To a great extent, the reason for such interest is due to
its importance in various branches of fundamental physics:
astrophysics, spectroscopy, solid state, and plasma physics. From
a physical point of view,  the first appearances of the influence
of a magnetic field $B$ on the atom are (i) changes in binding
energies, including the Zeeman level splitting which removes a
degeneracy; and (ii) the development of a nonvanishing quadrupole
moment $Q_{ab} \propto B_a B_b$ as a consequence of the
deformation of the spherical-symmetrical atomic shape. In contrast
to the former phenomenon, the latter has not been thoroughly
studied. Meanwhile, the appearance of a quadrupole moment leads
to a drastic change in the interaction of the atoms. A standard van
der Waals attraction which originates in the interaction of
induced dipoles is overtaken by 
quadrupole-quadrupole interaction
(which is repulsive when atoms are situated along magnetic line --
see Refs.\ \cite{Kl-Lozovik,Turbiner:1983}). In many applications (for
instance, for construction of an equation of state), one needs to
include the effects of atom-atom interactions. For example, a
study of pressure ionization of a strongly magnetized hydrogen
plasma was performed in Ref.\ \cite{PCS} with a simple occupation
probability model, which was based on a calculation of
quantum-mechanical atomic sizes \cite{P94}. This model is fully
adequate at sufficiently high temperatures $T$. However, in order
to extend the domain of applicability to lower $T$, where the neutral
fraction is large, electrical multipole interactions of atoms
should be taken into account. Therefore, quadrupole-quadrupole
interaction can be significant at certain plasma parameters.

For various quantum-mechanical states of the H atom in a magnetic
field, there have been accurate calculations of binding energies
\cite{Rosner,Kravchenko}, oscillator strengths \cite{Forster}, and
photoionization rates \cite{PPV}. Moreover, binding energies
\cite{P94,Vincke,P98}, geometrical sizes and oscillator strengths
\cite{P94,P98}, electric quadrupole transition probabilities
\cite{Cuv}, and photoionization cross sections \cite{PP97} have also
been successfully calculated for an atom {\em moving\/} in a
strong magnetic field (equivalent to an atom in crossed magnetic
and electric fields), which is an essentially three-dimensional
system. Despite this progress, up to now the quadrupole moment was
not studied basically with probably a single exception
\cite{Turbiner:1987}. A goal of the present Brief Report is to carry out
such a study for the ground state using (i) a variational method
and (ii) a method based on a solution of the Schr\"odinger
equation by expansion in Landau orbitals with coordinate-dependent
coefficients. We explore a range of magnetic field strengths $B$
from 0 to the ``relativistic" field $B_r\equiv\mel^2 c^3/(\hbar
e)=4.414\times 10^{13}$~G.

\section{Asymptotic results}

Hereafter, we will measure lengths in units of
$a_0\equiv\hbar^2/(\mel e^2)=0.529\,177\,${\AA} and energies in
units of ${\rm\,Ry}\equiv\frac12 e^2/a_0=13.6057$~eV. Assuming a
constant uniform magnetic field directed along the $z$-axis, we
take the vector potential ${\bf A}$ in the symmetric (axial)
gauge: $(A_x,A_y,A_z)=(B/2)\,(-y,x,0)$. A natural parameter of the
nonrelativistic theory is $\gamma=B/B_0$, where $B_0
 \equiv \mel^2 e^3/(\hbar^3 c)=2.3505\times10^9{\rm~G}$.
The field is called
``strong" if $\gamma\gtrsim1$.

Since the magnetic quantum number is equal to zero for the ground
state, the Hamiltonian has the form
\begin{equation}
\label{e.1}
 {\cal H} = - \Delta - \frac{2}{r} + \frac{\gamma^2}{4} \rho^2,
\quad
 \rho^2=x^2+y^2\ .
\end{equation}
Because of the axial symmetry of the problem, the components
$Q_{\alpha\beta}$ of the quadrupole tensor obey the following
relations (e.g., Ref.\ \cite{LaLi-Field}):
\bea &&
   Q_{xy}=Q_{yz}=Q_{zx}=0\ ,
\nonumber\\&&
\label{Qdiag}
  Q_{xx}=Q_{yy}=-\frac{1}{2}
  Q_{zz}=\langle z^2\rangle - \langle x^2\rangle\ .
\eea

In the weak-field limit, the usual perturbation theory gives
\cite{Turbiner:1987}
\beq
 -Q_{zz} = \frac{5}{2}
   \,\gamma^2-\frac{615}{32} \,\gamma^4+\ldots\ .
\label{Qzz-as0}
\eeq
 In the opposite case of an ultrastrong field, $\ln\gamma\gg1$,
when $\langle x^2\rangle\ll\langle z^2\rangle$, the longitudinal
motion can be separated, which gives rise to the one-dimensional
model \cite{Haines}. In the ground state, $\langle z^2\rangle$ is
mainly determined by the exponential tail of the one-dimensional
wave function, $\langle z^2\rangle \sim (2E)^{-1}$, where $E$ is
the binding energy. Using the method of Hasegawa and Howard
\cite{Hasegawa} for an evaluation of $E$, we find \beq
 - Q_{zz} \sim {1\over(\ln \gamma)^2}
      + {2\ln(\ln\gamma)\over(\ln\gamma)^3}
           + O\left({1\over(\ln\gamma)^3}\right)
\label{Qzz-as}
\eeq
at $\gamma\to\infty$.

\section{Variational method}
\label{sec-var}
In order to construct an adequate variational trial function
${\Psi_0}$, we follow a recipe formulated in Refs\ %
\cite{Turbiner:1980,Turbiner:1984,Turbiner:1989}. That is, the
potential $V_0 = (\Delta \Psi_0)/\Psi_0$ should reproduce the
Coulomb singularity at the origin, and the harmonic oscillator
behavior at large distances. Furthermore, the trial function
should have a correct functional expansion in coordinates at small
and large distances from the origin, as well as correct expansion
in powers of $B$. Since the ground state wave function has no
nodal surfaces in configuration space, we may write
${\Psi_0}=e^{-\phi}$, where ${\phi}$ is a smooth real function of
coordinates. The asymptotic behavior of this function was
calculated in Refs.\ \cite{Turbiner:1984a,Turbiner:1987}: 
\bea
 \phi &=&  
 \gamma \rho^2/4 + O(r)
 \quad (\rho\to\infty)\ ,
\\
 \phi &=& r + \gamma^2 \, (A r^3 + B r\rho^2+C r^2 +D \rho^2) 
\nonumber\\ && +
 O(\gamma^4 r^5)
 \quad (r\to0)\ , 
 \eea
 where $A,B,C$, and $D$ are known parameters. These expansions
prompt to choose the following seven-parametric trial function
\bea
 \Psi_{0}      &=&
 \exp\Big\{- \big[a^2 r^{2}+(\alpha_1 r^{3}
      + \alpha_2 \rho^2 r + \alpha_3 \rho^3 + \alpha_4
\rho r^{2})\,\gamma
\nonumber\\&&
      + (b_1 \rho^4  +
        \ b_2 \rho^2 r^2) \,\gamma^2/16 \,\big]^{1/2} \Big\}
\label{e.2}
\eea
 (cf.\ Refs.\ \cite{Turbiner:1987,Turbiner:1989}), where
$a, \alpha_{1-4}$, and $b_{1-2}$ are variational parameters. One can
check that the effective potential $V_0$ corresponding to this
trial function correctly reproduces the potential in \req{e.1} at
$r \to 0$ (Coulomb regime) and at $ \rho \to \infty$ (Landau
regime). Furthermore, \req{e.2} gives a correct functional form of
the first corrections in powers $B^2$ to the exponential phase of
the ground-state wave function (see Ref.\ \cite{Turbiner:1984a}) and,
even more importantly, the functional form of the first correction
to the Landau phase factor $\propto B \rho^2$ at large distances
(for a detailed discussion, see Ref.\ \cite{Turbiner:1987}). Thus,
\req{e.2} takes into account the available information on the
ground-state wave function of Hamiltonian (\ref{e.1}).

\section{Expansion in Landau orbitals}
\label{sec-Landau}
The shape of the atom is close to a sphere at $B\ll B_0$ and to a
cylinder at $B\gg B_0$. In the latter case, the expansion of the
atomic wave function over the Landau functions is appropriate
(e.g., Refs.\ \cite{P94,Rosner}).

If there were no Coulomb attraction, then the transverse part of
the wave function could be described by a Landau function
$\Phi_{ns}(\rho,\varphi)$ [where $\varphi$ is the polar angle in
the $(xy)$ plane], which satisfies the equation \beq
  - {1\over\rho}\,{\partial\over\partial\rho}\left(\rho\,
           {\partial\Phi_{ns}\over\partial\rho}\right)
     - {1\over\rho^2}\,{\partial^2\Phi_{ns}\over\partial\varphi^2}
            +{\gamma^2\over4}\,\Phi_{ns} = (2n+1)\,\gamma.
\label{Phi} \eeq
(e.g., Ref.\ \cite{ST}). Here $n$ is the Landau
quantum number and $s$ is the negative of the $z$ projection of
the electron orbital momentum ($n\geq 0$, $s\geq -n$). The Landau
functions form a complete orthogonal functional basis
on the $(xy)$ plane.

When the atom does not move as a whole across the field, $s$ is an
exact quantum number. Thus a wave function $\Psi$ can be
presented as

\begin{equation}
   \Psi({\bf r}) =
   \sum_{n} \Phi_{ns}(\rho,\phi) \, g_n(z).
\label{expan}
\end{equation}
The sum in \req{expan}, if truncated at some $n=N$, can be
considered as a variational trial function. The one-dimensional
functions $g_n$ are to be found numerically. The minimum of the
energy functional $\langle\Psi|{\cal H}|\Psi\rangle$ implies
zero functional derivatives: $\delta\langle\Psi|{\cal
H}|\Psi\rangle / \delta g_n(z) = 0$ ($\forall n$). Taking into
account \req{Phi}, one arrives at a system of coupled
differential equations for the set of $g_n(z)$ and $E$,
\beq
    {\dd^2\over\dd z^2}\,g_n(z)
 + 2 \sum_{n'} V_{nn'}^{(s)}(z)g_{n'}(z)
                 = (E + 2 n \gamma)\, g_n(z),
  \label{system}
\eeq
where
\beq
  V_{nn'}^{(s)}(z) =
 \int_0^\infty \!\!\!\rho\dd\rho \int_0^{2\pi} \!\!\!\dd\varphi\,
  \Phi_{ns}^*(\rho,\varphi)\,
  {1\over   r}
         \,\Phi_{n's}(\rho,\varphi).
\label{Vnn} \eeq 
The effective potentials [\req{Vnn}] can be
reduced to a finite sum of one-dimensional integrals feasible for
numerical calculation \cite{P94}.

Using the relations
\bea  &&  \left(\begin{array}{l} x^2 \\ y^2 \end{array}\right)
        = r_+ r_- \pm \frac12(r_+^2+r_-^2),
\\&&
  \sqrt{\gamma} r_+ \Phi_{ns} = \sqrt{n+s}\,\Phi_{n,s-1}
         - \sqrt{n+1}\,\Phi_{n+1,s-1},
\nonumber\\&&
  \sqrt{\gamma} r_- \Phi_{ns} = \sqrt{n+s+1}\,\Phi_{n,s+1}
         - \sqrt{n}\,\Phi_{n-1,s+1},
\nonumber \eea
where $r_\pm\equiv\rho\, e^{\pm i\varphi}$, one can
calculate the expectation values
\bea
   \langle z^2\rangle & = &
        \sum_{n\geq0} \int_{-\infty}^\infty z^2\,|g_n(z)|^2\dd z,
\label{z2}
\\
     {\langle x^2\rangle} & = & {\langle y^2\rangle} =
         \gamma\, \sum_{n\geq0} \int_{-\infty}^\infty \big[
    (2n+s+1)|g_n(z)|^2
\nonumber\\& - &
       2\,\sqrt{(n+1)(n+s+1)}\, |g_n^*(z) g_{n+1}(z)|\big]\dd z.
\label{x2}
\eea
 and finally the quadrupole moment $Q_{zz}$.

At $\gamma\gg1$ the first term $n=0$ dominates in the sum in
\req{expan}. Hence \req{x2} results in $\langle x^2
\rangle=\langle y^2 \rangle \approx (s+1)/\gamma$. It is
worthwhile to note that neglecting all terms in \req{expan}
except the one at $n=0$ is equivalent to the adiabatic
approximation used in early works (e.g., Refs.\ cite{CV77,Hasegawa}).

\section{Results and discussion}

The results of our calculations of the binding energy $E$ and the
quadrupole moment $Q_{zz}$ are presented in Table~\ref{tabQ}. When
available, we compare the binding energy with the most accurate
up-to-date results \cite{Kravchenko}.

The variational approach of Sec.~\ref{sec-var}, based on a {\it
single\/} seven-parametric function [\req{e.2}], gives very high
relative accuracy in the binding energy on the order of $10^{-7}$ at
small magnetic fields which then falls to $10^{-2}$ at the largest
studied magnetic fields.  Basically, this corresponds to the same
absolute accuracy (on the order of $10^{-7}$) in the {\em total\/}
energy for the whole explored range of magnetic fields. Two major
parameters $a$ and $b_1$ are changed as a function of magnetic
field in a very smooth and slow manner, from $a \sim 1$, $b_1 \sim
0.9$ for $10^9$~G to $a \sim 3$, $b_1 \sim 0.99$ at $10^{13}$~G,
respectively. Other parameters also vary smoothly and slowly.

For the second method (Sec.~\ref{sec-Landau}),
we retain $n,n'=0,1,\ldots,12$ in
the system of equations (\ref{system}) and solve it for the
ground state at $\gamma\geq1$ using the algorithm described in
Ref.~\cite{P94}. Then we calculate $Q_{zz}$ from \req{Qdiag}
using Eqs.~(\ref{z2}) and (\ref{x2}).

In Table~\ref{tabQ} we see that for the binding energy the method
of expansion in the Landau orbitals turns out to be more accurate
at $\gamma\gtrsim10$, whereas the variational method of
Sec.~\ref{sec-var} is superior at lower field strengths. This is
confirmed by a comparison with the results of
Ref.~\cite{Kravchenko}. We emphasize that our methods give very
close results for the quadrupole moment 
(the deviation does not exceed 10\%). This agrees with the
qualitative behavior found in \cite{Turbiner:1987}.

The data in Table~\ref{tabQ} can be approximated
by the expression
\bea
   -Q_{zz} &\approx& {\xi\,\gamma^{7/4} \over
            0.3392+(1+\xi^3)\,\gamma^{7/4}},
\label{fit}
\\&&
\mbox{where }
 \xi = 4 \ln(1 + 0.212 \,\gamma^{1/4}).
\nonumber
\eea
 This approximation reproduces the exact asymptotic behavior:
$-Q_{zz}\sim(\ln\gamma)^{-2}$ at $\gamma\to\infty$ and
$-Q_{zz}\sim\frac52\gamma^2$ at $\gamma\to0$. Its deviation from
the results in Table~\ref{tabQ} does not exceed a few percent in the
whole range of studied magnetic fields.

\begin{figure}
    \begin{center}
    \leavevmode
    \epsfxsize=86mm
    \epsfbox[30 270 570 590]{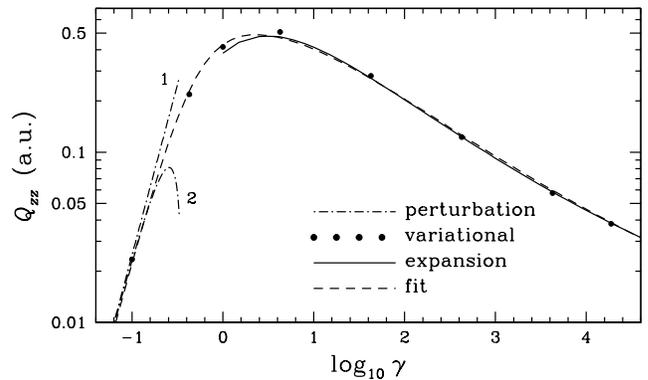}
    \end{center}
\caption{Absolute value of the quadrupole moment $Q_{zz}$ as a
 function of $\gamma=B/(2.35\times10^9$~G) calculated by the
 variational method (Sec.~\protect\ref{sec-var}) and the Landau 
 orbital expansion approach (Sec.~\protect\ref{sec-Landau}): 
 numerical results are compared with perturbation theory 
 [\protect\req{Qzz-as0}] [curve 1 corresponds to the first term 
 in Eq.~(\protect\ref{Qzz-as0}), curve 2 to two terms] and the
 analytic fit [\protect\req{fit}]. At $\gamma \rightarrow 0$ 
 the quadrupole moment $Q_{zz}$ tends to zero.}
 \label{fig}
\end{figure}

Figure \ref{fig} shows $|Q_{zz}|$ as a function of $\gamma$.
Numerical results obtained as described in Secs.~\ref{sec-var}
(shown by dots) and \ref{sec-Landau} (solid line) are
compared with perturbation theory of order $B^2$ and $B^4$
(lines marked `1' and `2', respectively) and with the fit
[\req{fit}] (dashed line). The quadrupole moment grows smoothly
with magnetic field increase, reaching the maximum at
$\gamma\approx3$ and then decreases. For the strongly elongated
atom at $\gamma\gg1$, the van der Waals constant can be roughly
estimated as $W\sim E\langle z^2\rangle^3 \propto
(\ln\gamma)^{-4}$. Thus $W$ decreases at $\gamma\to\infty$ at the
same rate as $Q_{zz}^2$. This means that the distance $R$, where
the van der Waals potential $\sim W/R^6$ becomes comparable with
the quadrupole-quadrupole interaction potential $\sim Q^2/R^5$,
tends to a finite value at $\gamma\to\infty$. Our results may have
an important impact on the modeling of relatively cool neutron
star atmospheres, whose spectra are being measured with the x-ray
telescopes onboard the recently launched {\em Chandra\/} and {\em
XMM-Newton\/} space observatories (e.g., Refs.\ \cite{XMM,Chandra}).

\acknowledgments

We are grateful to J.~Babb for reading the manuscript and useful
comments. A.~T.\ thanks J.~C.~Lopez Vieyra for help with numerical
calculations. A.~P.\ thanks M.~V.~Ivanov for 
a valuable remark and acknowledges support from RFBR (Grant
No.\ 99-02-18099). The work of A.~T. is supported in part by DGAPA
Grant No.\ IN105296 (M\'exico).

\pagebreak

\begin{table}
\caption{Binding energy $E$ and absolute value of
the quadrupole moment $Q_{zz}$ at
different magnetic fields $B$ calculated (a) by the variational
method and (b) by expansion in the Landau basis. Rounded-off data
from Ref.~\protect\cite{Kravchenko} are given for comparison. }
\label{tabQ}
\begin{tabular}{llllll}
$B$ & \multicolumn{3}{c}{$E$ (Ry)} & \multicolumn{2}{c}{$-Q_{zz}$ (a.u.)}\\
 & \mc{(a)} & \mc{(b)} & \mc{Ref.~\cite{Kravchenko}}& \mc{(a)} & \mc{(b)} \\
\noalign{\smallskip}\hline\noalign{\smallskip}
 $0.1\,B_0$ & 1.095\,05274  & \mc{--} & 1.095\,05296 
                                          & 0.0235 & \mc{--} \\
 $10^9$~G          & 1.346\,292 & \mc{--} & & 0.2185 & \mc{--} \\
 $B_0$ & 1.662\,322  & 1.63 & 1.662\,338 
                                             & 0.4155 & 0.38 \\
 $10^{10}$~G        & 2.614\,73 & 2.61    & & 0.5085 & 0.48 \\
 $10\,B_0$ & 3.4948 & 3.490 & 3.495\,6 
                                             & 0.4370 & 0.447 \\
 $10^{11}$~G        & 5.713     & 5.717   & & 0.2806 & 0.290 \\
 $100\,B_0$& 7.5642 & 7.579 & 7.579\,6 
                                             & 0.2071 & 0.217 \\
 $10^{12}$~G        & 11.87     & 11.924   & & 0.1228 & 0.1308 \\
 $1000\,B_0$& 15.23 & 15.325 & 15.324\,9 
                                             & 0.0915 & 0.0981 \\
 $10^{13}$~G        & 22.5      & 22.77   & & 0.0576 & 0.0620 \\
 $B_r$ & 32.5    & 32.92   & & 0.0380 & 0.0406 \\
\end{tabular}
\end{table}


\begin{references}

\bibitem{CV77}
V.~Canuto and J.~Ventura,
{Fundam.\ Cosm.\ Phys.} {\bf 2}, 203 (1977).

\bibitem{Garstang:1977}
 R.~H.~Garstang,  {Rep.\ Prog.\ Phys.} {\bf 40}, 105 (1977).

\bibitem{Ruder:1994}
        H.~Ruder, G.~Wunner, H.~Herold, and F.~Geyer,
{\it Atoms in Strong Magnetic Fields\/} (Springer, Berlin, 1994).

\bibitem{Kl-Lozovik}
        Yu.~E.~Lozovik and A.~V.~Klyuchnik,
        {Phys. Lett.} {\bf66A}, 282 (1978).

\bibitem{Turbiner:1983}
        A.~V.~Turbiner,
           {Pisma Zh.\ Eksp.\ Teor.\ Fiz.} {\bf 38}, 510 (1983)
[JETP Lett. {\bf 38}, 618 (1983)].

\bibitem{PCS}
        A.~Y. Potekhin, G.~Chabrier, and Yu.~A. Shibanov,
  Phys.\ Rev. E {\bf 60}, 2193 (1999); {\bf 63}, 019901(E) (2001)

\bibitem{P94}
         A.~Y. Potekhin,
\JPB{27}, 1073 (1994).

\bibitem{Rosner}
W.~R\"osner, G.~Wunner, H.~Herold, and H.~Ruder,
{J.\ Phys.\ B: At.\ Mol.\ Phys.} {\bf 17}, 29 (1984).

\bibitem{Kravchenko}
        Yu.~P. Kravchenko, M.~A. Liberman, and B.~Johansson,
{Phys.\ Rev. A} {\bf 54}, 287 (1996).

\bibitem{Forster}
H.~Forster, W.~Strupat, W.~R\"{o}sner, G.~Wunner, H.~Ruder,
and H.~Herold,
\JPB {\bf 17} 1301 (1984).

\bibitem{PPV}
A.~Y.~Potekhin, G.~G.~Pavlov, and J.~Ventura,
{Astron.~Astrophys.}, {\bf 317}, 618 (1997).

\bibitem{Vincke}
M.~Vincke, M.~Le Dourneuf, and D.~Baye, \JPB{25}, 2787 (1992).

\bibitem{P98}
A.~Y.~Potekhin,
\JPB{31}, 49 (1998).

\bibitem{Cuv}
C.~Cuvelliez, D.~Baye, and M.~Vincke, {Phys.\ Rev. A} {\bf 46},
4055 (1992).

\bibitem{PP97}
A.~Y.~Potekhin and G.~G.~Pavlov, 
{Astrophys.\ J.} {483}, 414 (1997).

\bibitem{Turbiner:1987}
        A.~V.~Turbiner,
           {Yad.\ Fiz.} {\bf 46}, 204 (1987)
  [{Sov.\ J.\ Nucl.\ Phys.} {\bf 46}, 125 (1987)].

\bibitem{LaLi-Field}
L.~D. Landau and E.~M. Lifshitz,
{\it The Classical Theory of Fields\/} (Pergamon, Oxford, 1981).

\bibitem{Haines}
 L.~K. Haines and D.~H. Roberts,
 {Am.\ J.\ Phys.} {\bf 37}, 1145 (1969).

\bibitem{Hasegawa}
H.~Hasegawa and R.~E.~Howard,
{J.\ Phys.\ Chem.\ Solids} {\bf 21}, 179 (1961).

\bibitem{Turbiner:1980}
        A.~V.~Turbiner,
           {Zh.\ Eksp.\ Teor.\ Fiz.} {\bf 79}, 1719 (1980)
   [{Sov.\ Phys. JETP} {\bf 52}, 868 (1980)].

\bibitem{Turbiner:1984}
        A.~V.~Turbiner,
           {Usp.\ Fiz.\ Nauk} {\bf 144}, 35 (1984)
[{Sov. Phys. Usp. \bf 27}, 668 (1984).

\bibitem{Turbiner:1989}
        A.~V.~Turbiner,
     Doctor of Sciences Thesis, ITEP, Moscow, 1989 (unpublished)

\bibitem{Turbiner:1984a}
        A.~V.~Turbiner,
           {J.\ Phys. A} {\bf 17}, 859 (1984).

\bibitem{ST}
A.~A. Sokolov and I.~M. Ternov,
{\it Synchrotron Radiation\/} (Academic, Berlin, 1968).

\bibitem{XMM}
F. Paerels, K. Mori, C. Motch, F. Haberl, V. E. Zavlin,
S. Zane, G. Ramsay, M. Cropper, and B. Brinkman, 
Astron.\ Astrophys. 365, L298 (2001).

\bibitem{Chandra}
V. E. Zavlin, G. G. Pavlov, D. Sanwal, and J. Tr\"umper,
Astrophys.\ J. {\bf 540}, L25 (2000).

\end{references}
\end{document}